\documentstyle[12pt,epsfig]{article}  
\begin{document} 
\baselineskip=20pt

%%%%%%%%%%%%%%%%%%%%%%%% NEW DEFINITIONS
\def\la{\mathrel{\mathpalette\fun <}}
\def\ga{\mathrel{\mathpalette\fun >}}
\def\fun#1#2{\lower3.6pt\vbox{\baselineskip0pt\lineskip.9pt
\ialign{$\mathsurround=0pt#1\hfil##\hfil$\crcr#2\crcr\sim\crcr}}} 
%%%%%%%%%%%%%%%%%%%%%%%% TITLE PAGE

\begin{titlepage} 
\begin{center}
{\Large \bf Medium-modified fragmentation of b-jets tagged by a leading muon 
in ultrarelativistic heavy ion collisions} \\

\vspace{4mm}

I.P.~Lokhtin$^a$, 
L.I.~Sarycheva$^b$,  
A.M.~Snigirev$^c$, 
K.Yu.~Teplov$^d$  \\
M.V. Lomonosov Moscow State University, D.V. Skobeltsyn Institute of Nuclear 
Physics, \\
119992, Vorobievy Gory, Moscow, Russia 
\end{center}  

\begin{abstract}  
The possibility to observe the medium-modified fragmentation of hard b quarks tagged
by a leading muon in ultrarelativistic heavy ion collisions is analyzed. 
We have found that reasonable statistics, $\sim 10^4$ events per 1 month of 
LHC run with lead beams, can be expected for the realistic geometrical acceptance 
and kinematic cuts. The numerical estimates on the effect of the medium-induced 
softening b-jet fragmentation function are given. 
\end{abstract}

\bigskip

\vspace{100mm}
\noindent
---------------------------------------------\\
$^a$ e-mail: igor@lav01.sinp.msu.ru\\
$^b$ e-mail: lis@alex.sinp.msu.ru\\
$^c$ e-mail: snigirev@lav01.sinp.msu.ru\\
$^d$ e-mail: teplov@lav01.sinp.msu.ru \\ 
\end{titlepage}   

\section{Introduction} 
The experimental investigation of ultrarelativistic nuclear collisions offers 
a unique possibility of studying the properties of strongly interacting matter 
at a high energy density. In that regime the hadronic matter is expected to become 
deconfined and a gas of asymptotically free quarks and gluons is formed. This is
a quark-gluon plasma (QGP), in which the colour interactions between the  
partons are screened owing to collective effects~\cite{qm}. One of the important 
tools to study QGP properties in heavy ion collisions is a QCD jet production. 
Medium-induced energy loss of energetic partons, the so-called jet quenching, 
has been proposed to be very different in cold nuclear matter and in QGP, 
resulting in many challenging observable phenomena~\cite{baier_rev}. Recent RHIC 
data on suppression of inclusive high-p$_T$ charge and neutral hadron production 
from STAR~\cite{star}, PHENIX~\cite{phenix}, PHOBOS~\cite{phobos} and 
BRAHMS~\cite{brahms} are in agreement with the jet quenching 
hypothesis~\cite{Wang:2004}. However direct event-by-event reconstruction of 
jets and their characteristics is not available in RHIC experiments at the 
moment, while the assumption that the integrated yield of all high-$p_T$ particles 
originates only from the jet fragmentation is not obvious.

At LHC a new regime of heavy ion physics will be reached at 
$\sqrt{s_{\rm NN}}=5.5$ TeV where hard and semi-hard QCD multi-particle 
production can dominate over underlying soft events. The initial gluon densities 
in Pb$-$Pb reactions at LHC are expected to be significantly higher than at 
RHIC, implying a stronger partonic energy loss which can be observable in various 
new channels~\cite{lhc-jets,lhc-hq,lhc-phot}. In particular, the influence of
the medium-modified fragmentation of heavy quarks on dilepton spectra was analyzed 
in~\cite{kamp98,lin98,lokhtin01}. Since the estimated event rates for b quark
production at LHC energies are expected to be high enough, in combination with 
high-$p_T$ jet production by gluon and light quark fragmentation this can give 
important information about the medium-induced effects for both light and heavy 
partons in nucleus-nucleus interactions at the LHC.

In previous paper~\cite{phl03} we analyzed the possibility to observe the  
medium-induced softening jet fragmentation function (JFF) of light partons 
tagged by a leading
neutral or charge hadron in heavy ion collisions at the LHC. In this paper the
possibility to measure the medium-modified b-JFF by a leading muon is suggested and
analyzed for LHC conditions. In Sect.~2 we give the main definitions of JFF, 
calculate the cross section of $B(\rightarrow {\rm leading} ~\mu  )$ production at LHC 
energies with PYTHIA generator and estimate the expected event rate for the realistic 
geometrical acceptance and kinematic cuts. Sect.~3 describes shortly a model 
of partonic energy loss in QGP used to evaluate the sensitivity of b-JFF to the  
jet quenching effect. Discussion on numerical results and summary can be found 
in Sect.~4.   

\section{$B(\rightarrow {\rm leading}~\mu)$ production at LHC}
Let us recall that the jet fragmentation function $D(z)$ determines the probability 
for a final ``jet-induced'' particle to carry a fraction $z$ of the jet transverse
momentum $p_T^{\rm jet}$. In nuclear $AA$ interactions JFF for leading
particles can be defined as~\cite{phl03}: 
\begin{equation}
\label{dz}
D(z)=\int \limits_{z\cdot p_{T~{\rm min}}^{\rm jet}} d(p^L_T)^2 dy dz' 
\frac{dN_{AA}^{\rm h(k)}}{d(p^L_T)^2dydz'} \delta \left( z-\frac{p^L_T}{p^{\rm
jet}_T} \right) \Bigg/ \int \limits_{p_{T~{\rm
min}}^{\rm jet}} d(p_T^{\rm jet})^2 dy 
\frac{dN_{AA}^{\rm jet(k)}}{d(p_T^{\rm jet})^2dy}~, 
\end{equation}
where $p^L_T \equiv z p_T^{\rm jet}=z'p_T$ is the leading particle transverse 
momentum, $z'$ is the momentum fraction relatively to $p_T$ of the parent parton
(of course, without energy loss $z=z'$ in the leading order of perturbative
QCD),
$p_{T~{\rm min}}^{\rm jet}$ is the minimum threshold for energy of 
observable jets. The rate of $k$-type jets in mid-rapidity with transverse 
momentum $p_T$ in $AA$ collisions at the given impact parameter $b$ is estimated as 
\begin{equation}
\label{vertex}
\frac{dN_{AA}^{\rm jet(k)}}{d(p_T^{\rm jet})^2dy} (b) = \int\limits_0^{2\pi} d \psi 
\int\limits_0^{r_{max}}r dr T_A(r_1) T_A(r_2)   
\frac{d\sigma^{\rm jet(k)}(p_T^{\rm jet}+\Delta p_T^{\rm jet}(r, \psi, \theta_0))}{dp_T^2dy} , 
\end{equation} 
and is determined by the absolute value of partonic energy loss as well as by the 
angular radiation spectrum. Here $r_{1,2} (b,r,\psi)$ are the distances between 
the nucleus centers and the 
jet production vertex $V(r\cos{\psi}, r\sin{\psi})$; $r_{max} (b, \psi) \le R_A$ 
is the maximum possible transverse distance $r$ from the nuclear collision axis 
to the $V$; $R_A$ is the radius of the nucleus $A$; $T_A(r_{1,2})$ is the nuclear 
thickness function (see Ref.~\cite{lokhtin00} for detailed nuclear geometry 
explanations). The effective shift $\Delta p_T^{\rm jet} (r, \psi, \theta_0)$ 
of the jet momentum spectrum depends on the jet angular cone size $\theta_0$ (see
Fig.1). In the leading order of perturbative QCD the jet production cross section, 
$d\sigma^{\rm jet(k)}/(dp_T^2dy)$, is calculated in our case with 
PYTHIA6.2~\cite{pythia}. The rate of high-$p_T$ jet-induced hadrons is estimated
as 
\begin{equation}
\frac{dN_{AA}^{\rm h(k)}}{d(p_T^L)^2dydz'} (b) = \int\limits_0^{2\pi} d \psi 
\int\limits_0^{r_{max}}r dr T_A(r_1) T_A(r_2) 
\frac{d\sigma^{\rm jet(k)}(p_T+\Delta p_T(r, \psi))}{dp_T^2dy} \frac{1}{z'^2}   
D^h_k(z', p_T^2)~,   
\end{equation} 
where the shift $\Delta p_T$ of the hadron momentum distribution generally is not equal 
to the mean in-medium partonic energy loss due to the steep fall-off of the 
$p_T$-spectrum~\cite{Baier:2001}.

For jets initiated by light hadrons, the leading particles are the charged or neutral
hadrons. However for heavy quark initiated jets there is the possibility to have
a leading muon produced by semileptonic meson decays. Thus jet tagged by high-$p_T$
muon can be identified as a heavy quark jet. Note that $\approx 20 \% $ of 
$B-$mesons and $\approx 12 \%$ of $D-$mesons decay to muons, about half of the muons 
from $B-$decays being produced through an intermediate $D$~\cite{part_data}. 

We used PYTHIA6.2~\cite{pythia} with CTEQ5L pdf parameterization to calculate the
cross section of b-jet production and the corresponding spectra at 
$\sqrt{s_{\rm pp}}=5.5$ TeV and to estimate the expected event rate for the  
realistic geometrical acceptance and kinematic cuts. To be specific, the 
geometry of Compact Muon Solenoid (CMS) detector is 
considered~\cite{cms94,note00-060}: the pseudo-rapidity coverage 
$\mid \eta \mid < 3$ for jets and $\mid \eta \mid < 2.4$ for muons. We define
the muon as a leading particle if it belongs a hard jet and carries larger 
$20\%$ of the jet transverse momentum. To be specific, the jet energy is determined 
here as the total transverse energy of the final particles collected around the direction 
of a leading particle inside the cone $R=\sqrt{\Delta \eta ^2+\Delta \varphi ^2}=0.5$, 
where $\eta$ and $\varphi$ are the pseudorapidity and the azimuthal angle 
respectively. Extra cuts $p_T^{\mu }> 5$ GeV/$c$ and $E_T^{\rm jet}>50$ GeV were 
applied. Then the corresponding $pp$ cross section for $B (\rightarrow ~{\rm
leading}~\mu)$ production is $\approx 0.7~$pb, and Pb$-$Pb cross section is 
estimated as $0.7~{\rm pb} \times (207)^2 \approx 0.03~$mb. The corresponding 
event rate in a one month Pb$-$Pb run (assuming 15 days of data taking), 
$H=1.3\times 10^6$ s, with luminosity $L =5 \times 10^{26}~$cm$^{-2}$s$^{-1}$, 
is $N_{\rm ev}= H \sigma_{PbPb} L \approx 2 \times 10^4$ in this case. 
Increasing the minimal jet energy results in reducing expected statistics, e.g. for 
$E_T^{\rm jet}>100$ GeV the estimated rate is only $\sim 10^3$ events.  

\section{The model for simulation of jet quenching}
In order to  test the sensitivity of b-JFF to the jet quenching, the following 
event-by-event Monte Carlo simulation procedure was applied (see
Refs.~\cite{lokhtin00,pyquen} for details of the model). 

\begin{itemize} 
\item Generation of the initial parton spectra with PYTHIA (fragmentation {\em off}). 
\item Generation of the jet production vertex at the impact parameter 
$b$ according to the distribution
\begin{equation} 
\frac{dN^{\rm jet}}{d\psi dr} (b) = \frac{T_A(r_1) T_A(r_2)}
{\int\limits_0^{2\pi} d \psi \int\limits_0^{r_{max}}r dr T_A(r_1) T_A(r_2)}~.
\end{equation} 
\item Calculation of the cross section $\sigma = \int dt d\sigma/dt$ for scattering of 
a parton with energy $E$ off the ``thermal'' partons with energy (or effective 
mass) $m_0 \sim 3T \ll E$ ($T$ is the medium temperature) and generation of
the transverse momentum transfer $t_i$ according to the distribution
\begin{equation} 
\label{sigt} 
\frac{d\sigma }{dt} \cong C \frac{2\pi\alpha_s^2(t)}{t^2} \frac{E^2}{E^2-m_q^2}~,~~~~
\alpha_s = \frac{12\pi}{(33-2N_f)\ln{(t/\Lambda_{QCD}^2)}} \>~.
\end{equation} 
Here $C = 9/4, 1, 4/9$ for $gg$, $gq$ and $qq$ scatterings respectively, 
$\alpha_s$ is the QCD running coupling constant for $N_f$ active quark flavors, 
and $\Lambda_{QCD}$ is the QCD scale parameter which is of the order of the 
critical temperature, $\Lambda_{QCD}\simeq T_c \simeq 200$ MeV. The integrated 
cross section $\sigma$ is regularized by the Debye screening mass squared 
$\mu_D^2 (T) \simeq 4\pi \alpha _s T^2(1+N_f/6)$. The maximum momentum transfer 
$t_{\rm max}=[ s-(m_q+m_0)^2] [ s-(m_q-m_0)^2 ] / s$ where $s=2m_0E+m_0^2+m_q^2$, 
$m_q$ is the hard parton mass.
\item Generation of the transverse distance between scatterings, 
$l_i = (\tau_{i+1} - \tau_i)p_T/E$:  
\begin{equation} 
\frac{dP}{dl_i} = \lambda^{-1}(\tau_{i+1}) \exp{(-\int\limits_0^{l_i}
\lambda^{-1} (\tau_i + s)ds)} ~,~~ \lambda^{-1}(\tau ) =\sigma (\tau ) \rho 
(\tau )~,
\end{equation}  
where $\tau$ is the proper time, $\lambda$ is the in-medium mean free path, $\rho 
\propto T^3$ is the medium density. 
\item Reducing the parton energy by collisional and radiative loss per scattering $i$:
\begin{equation} 
\Delta E_{{\rm tot},i} = \Delta E_{{\rm col},i} + \Delta E_{{\rm rad},i}~,
\end{equation} 
where the collisional part is calculated in the high-momentum transfer approximation, 
\begin{equation} 
\label{col} 
\Delta E_{{\rm col},i} = \frac{t_i}{2 m_0}~,
\end{equation} 
and the radiative part is generated according to the energy spectrum $dI/d\omega $ 
obtained in the frame of BDMS model~\cite{baier} generalized to the case of 
heavy quarks -- the ``dead cone'' approximation~\cite{dc} (but note there are exist 
more recent developments on heavy quark energy loss in the 
literature~\cite{djor,armesto}):  
\begin{eqnarray}
\label{radmass} 
\frac{dI}{d\omega }| _{m_q \ne 0} =  \frac{1}{(1+(l\omega )^{3/2})^2}
\frac{dI}{d\omega }| _{m_q=0}~, ~~~ l=\left( \frac{\lambda}{\mu_D^2}\right) ^{1/3}
\left( \frac{m_q}{E}\right) ^{4/3}~,\\ 
\label{radiat}
\frac{dI}{d\omega }| _{m_q=0} = \frac{2 \alpha_s (\mu_D^2) \lambda C_R}{\pi L \omega }
\left[ 1 - y + \frac{y^2}{2} \right] 
\>\ln{\left| \cos{(\omega_1\tau_1)} \right|} 
\>, \\  
\omega_1 = \sqrt{i \left( 1 - y + \frac{C_R}{3}y^2 \right)   
\bar{\kappa}\ln{\frac{16}{\bar{\kappa}}}}
\quad \mbox{with}\quad 
\bar{\kappa} = \frac{\mu_D^2\lambda_g  }{\omega(1-y)} ~, 
\end{eqnarray}
where $\tau_1=L/(2\lambda_g)$, $y=\omega/E$ is the fraction of the hard parton 
energy carried by the radiated gluon, and $C_R = 4/3$ is the quark color factor. 
A similar expression for the gluon jet can be obtained by substituting 
$C_R=3$ and a proper change of the factor in the square bracket in (\ref{radiat}), see
Ref.~\cite{baier}. The allowed range of values 
$\omega _i= \Delta E_{{\rm rad},i}$ in (\ref{radiat}) is from 
$\omega_{\min}=E_{LPM}=\mu_D^2\lambda_g$, the minimal radiated gluon energy in 
the coherent LPM regime, to initial jet energy $E$. 
\item Calculation of the parton transverse momentum kick due to elastic scattering 
$i$:
\begin{equation} 
\Delta k_{t,i}^2 =(E-\frac{t_i}{2m_{0i}})^2-(p-\frac{E}{p}\frac{t_i}{2m_{0i}}-
\frac{t_i}{2p})^2-m_q^2 .
\end{equation} 
\item Formation of the additional (in-medium emitted) gluon with the energy 
$\omega _i=\Delta E_{{\rm rad},i}$ and the direction relatively to the parent parton 
determined according to one of two possible simple parameterizations for 
the emission angle $\theta$: the ``small-angular'' parameterization,  
\begin{equation} 
\label{sar} 
\frac{dN^g}{d\theta}\propto \sin{\theta} \exp{\left( -\frac{(\theta-\theta
_0)^2}{2\theta_0^2}\right) }~, 
\end{equation}
where $\theta_0 \sim 5^0$ is the typical angle of the coherent gluon radiation 
estimated in~\cite{lokhtin98}; or the ``wide-angular'' parameterization, 
\begin{equation} 
\label{war} 
\frac{dN^g}{d\theta}\propto \frac{1}{\theta}~.  
\end{equation}
\item Halting the parton rescattering if {\em 1)} a parton escapes from the 
dense zone, or {\em 2)} QGP cools down to $T_c=200$ MeV, or {\em 3)} a parton 
loses so much energy that its $p_T (\tau)$ drops below $2T (\tau)$. 
\item In the end of each event adding new (in-medium emitted) gluons into 
PYTHIA parton list and rearrangements of partons to update string formation are
performed.  
\item Formation of the final particles by PYTHIA (fragmentation {\em on}).  
\end{itemize} 

The medium was treated as a boost-invariant longitudinally expanding quark-gluon 
fluid, and partons as being produced on a hyper-surface of equal proper times 
$\tau$~\cite{bjorken}. In order to simplify numerical calculations in the 
original version of the model we omit the transverse expansion and viscosity of 
the fluid using the well-known scaling solution due to Bjorken~\cite{bjorken} 
for a temperature and density of QGP at $T > T_c \simeq 200$ MeV:
\begin{equation}
\varepsilon(\tau) \tau^{4/3} = \varepsilon_0 \tau_0^{4/3},~~~~
T(\tau) \tau^{1/3} = T_0 \tau_0^{1/3},~~~~ \rho(\tau) \tau = \rho_0 \tau_0 .
\end{equation}
For certainty we used the initial conditions for the gluon-dominated plasma 
formation expected for central Pb$-$Pb collisions at LHC~\cite{esk}: 
$$\tau_0 \simeq 0.1~{\rm fm/c}, ~~~~T_0 \simeq 1~{\rm GeV}, ~~~~\rho_g 
\approx 1.95T^3 ~.$$ 

\newpage

\section{Numerical results and conclusions}

Figure 2 shows the b-JFF~(\ref{dz}) of leading muons for the cases without and 
with medium-induced energy loss in central Pb$-$Pb collisions with the two  
parameterizations of the distribution on the gluon emission angles (\ref{sar}) and 
(\ref{war}); the same geometrical acceptance and kinematic cuts as described in 
Sect.~2 were used. One can see the softening b-JFF due to partonic 
energy loss at $z \ga 0.4$. The effect enhances with $z$ decreasing (see Fig.3) and is more 
pronounced for the small-angular radiation. The reason for the latter fact is
following. The contribution of the small-angular radiation to the total jet 
energy loss (due to ``out-of-cone'' partonic energy loss) is much lesser as compared 
with the broad-angular radiation. The former does not disappear totally mostly 
because 
not only leading (parent) parton, but all partons of a jet pass through the 
dense medium and emit gluons under the angles $\theta$ relatively to their proper 
directions, which in general may not coincide with the jet axis (determined by 
the direction of a leading particle) and sometimes be even at the jet periphery. 
The broad-angular radiation increases the ``out-of-cone'' part of partonic energy 
loss and thus decreases the final jet transverse momentum $p_T^{\rm jet}$ (which is 
the denominator in the definition of $z\equiv p_T^L/p_T^{\rm jet}$ in 
JFF~(\ref{dz})) without any influence on the numerator of $z$ and, as a 
consequence, in reducing effect on JFF softening.  

Note that in the real experimental situation the jet observables will be 
sensitive to the accuracy of jet energy reconstruction in a high multiplicity 
environment, in particular, to the systematic jet energy loss. However, since 
the average reconstructed jet energy in Pb$-$Pb collisions is expected to be 
the same as in $pp$ interactions (see section ``Jet detection at CMS'' in
Ref.~\cite{lhc-jets}), the short measure of jet energy will be the well-controlled  
systematic error for heavy ion as well for $pp$ collisions, and it can be 
taken into account using the standard calibration procedure.

In summary, the channel with the muon tagged b-jet production in ultrarelativistic 
heavy ion collisions was first analyzed. The reasonable statistics, $\sim 10^4$ 
events per 1 month of LHC run with lead beams, can be expected for the realistic 
geometrical acceptance and kinematic cuts. The effect on the medium-modified b-jet 
fragmentation was numerically studied for Pb$-$Pb collisions at the LHC.
The significant softening b-jet fragmentation function determined by the 
absolute value of partonic energy loss and the angular radiation spectrum is 
predicted. 
 
\bigskip

{\it Acknowledgments}.  Discussions with Yu.L.~Dokshitzer, O.L.~Kodolova,
C.~Roland, I.N.~Vardanyan, R.~Vogt, B.~Wyslouch, B.G.~Zakharov and 
G.M.~Zinovjev are gratefully acknowledged. This work is supported by grant 
N 04-02-16333 of Russian Foundation for Basic Research.

\begin{figure}[hbtp] 
\begin{center} 
\makebox{\epsfig{file=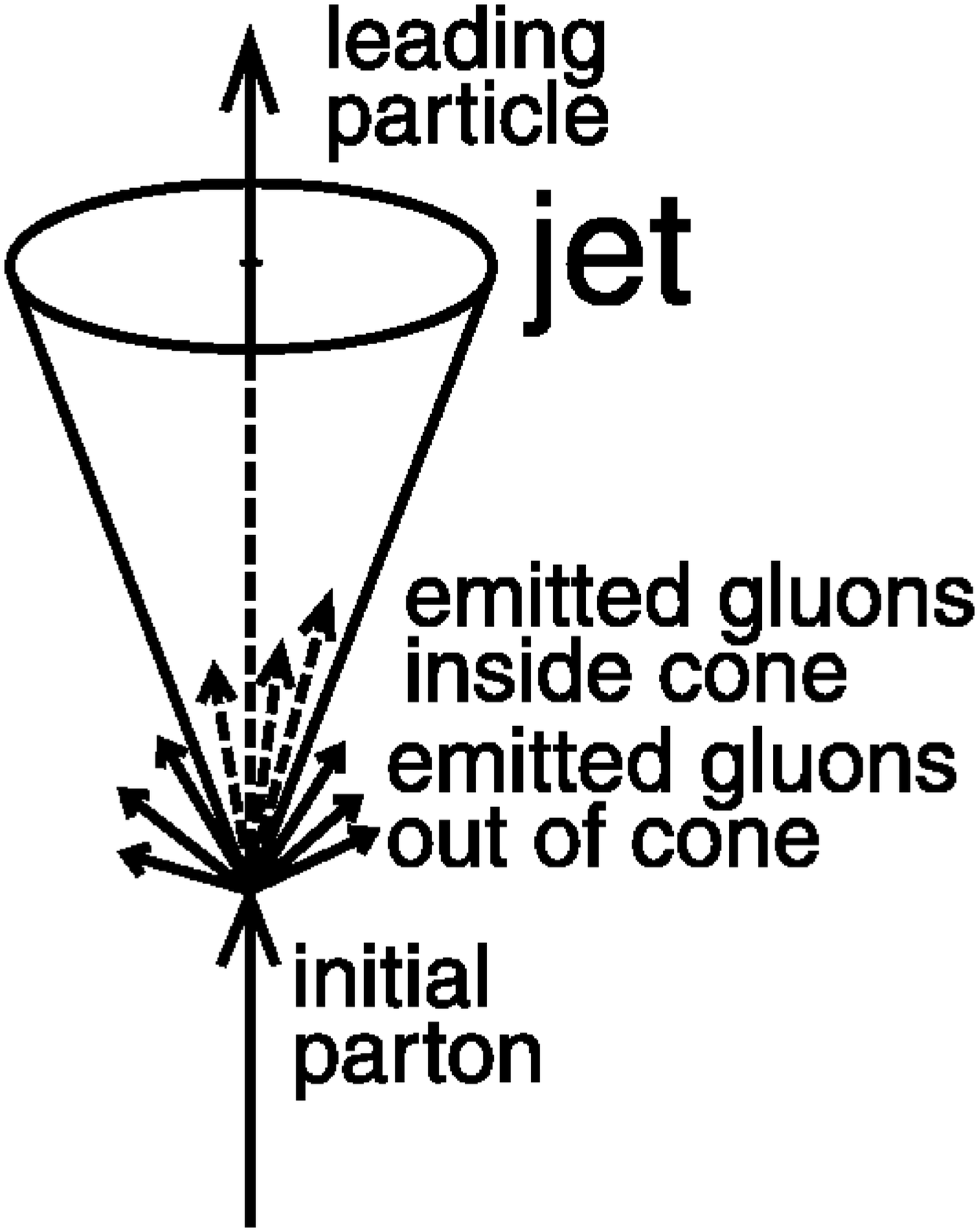, height=170mm}}   
\vskip 1cm 
\caption{\small  The schematic view of a jet with gluons emitted inside and
outside jet cone.} 
\end{center}
\end{figure}

\begin{figure}[hbtp] 
\begin{center} 
\makebox{\epsfig{file=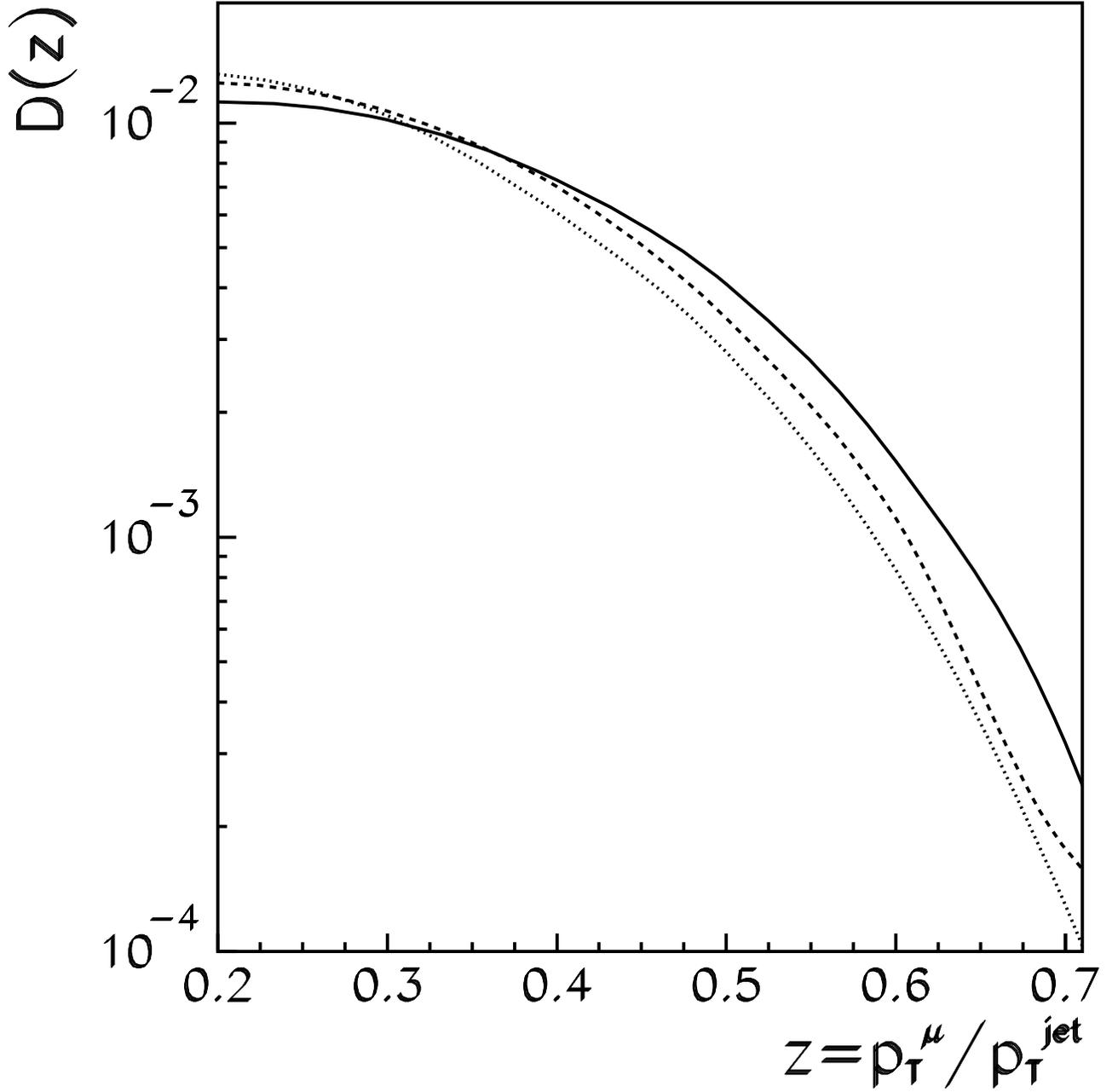, height=170mm}}   
\vskip 1cm 
\caption{\small  B-jet fragmentation function for leading muons without (solid   
curve) and with medium-induced partonic energy loss for the 
``small-angular'' (\ref{sar}) (dotted curve) and the ``broad-angular'' 
(\ref{war}) (dashed curve) parameterizations of emitted gluon spectrum in 
central Pb$-$Pb collisions. Applied kinematic cuts are described in 
the text.}  
\end{center}
\end{figure}

\begin{figure}[hbtp] 
\begin{center} 
\makebox{\epsfig{file=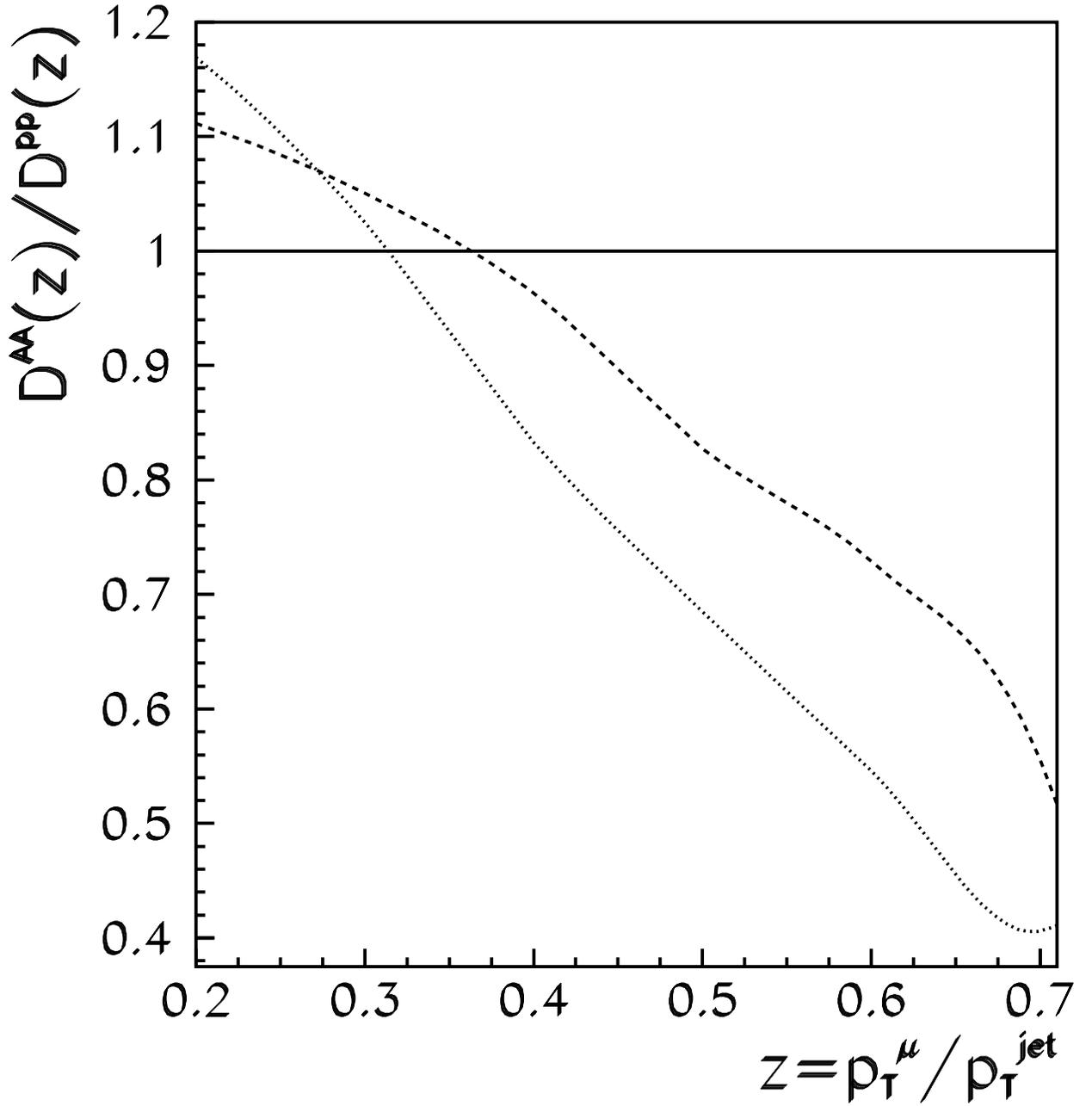, height=170mm}}   
\vskip 1cm 
\caption{\small  The ratio of B-jet fragmentation function for leading muons with
energy loss to one without energy loss in central Pb$-$Pb collisions. 
The dotted curve is the result for the ``small-angular'' radiation, the dashed 
curve - for the ``broad-angular'' radiation. Applied kinematics cuts are 
described in the text.}  
\end{center}
\end{figure}

\end{document}